\documentstyle[aps,prl,epsf]{revtex} 

\begin{document}
\twocolumn[{
\draft

\title{
Long-Tailed Trapping Times and L\'evy Flights  \\
in a Self-Organized Critical Granular System
}
\author{ 
Mari\'an Bogu\~n\'a\cite{email.mbogunya}  
and 
\'{A}lvaro Corral\cite{email.alvaro}
}
\address{ 
Departament de F\'{\i}sica Fonamental, 	
Universitat de Barcelona, Diagonal 647, E-08028 
Barcelona, Spain \\
}
\date{\today}
\maketitle  
\widetext
\begin{abstract}
\leftskip 54.8 pt
\rightskip 54.8 pt
We present a continuous time random walk model for the scale-invariant 
transport found in a self-organized  critical rice pile
[Christensen {\it et al.}, Phys. Rev. Lett. {\bf 77}, 107 (1996)].
From our analytical results it is shown that
the dynamics of the experiment can be explained in terms of 
L\'evy flights for the grains and
a long-tailed distribution of trapping times.
Scaling relations for the exponents of these distributions
are obtained.
The predicted microscopic behavior is confirmed
by means of a cellular automaton model.
\end{abstract}
\leftskip 54.8 pt

}]
\narrowtext

Self-organized criticality (SOC), or the spontaneous
emergence of scale invariance in nonequilibrium systems
has attracted a great interest 
as an explanation of fractal behavior in nature \cite{Bak87}.
Sandpiles rapidly became the paradigm of SOC,
but it has been only very recently that they have been shown
to be characterized by power-law distributions of avalanches
\cite{Frette96,Feder95}.
On the other hand, understanding the complex behavior of granular
media is a challenge of fundamental 
physics {\it per se}.
Between many other amazing properties, granular systems
can behave simultaneously as solid or liquids
and show a glassy dynamics with extremely slow relaxations \cite{Jaeger}.
In addition, the transport properties found in granular systems 
displaying SOC \cite{Christensen96} turns out to be very similar 
to the dispersive transport
taking place in amorphous semiconductors and polymers \cite{Scher}.
Finally, there are close connections between sandpile models and interface 
depinning \cite{Paczuski}.

The experimental system that we want to analyze is a rice pile,
built in the narrow gap between two vertical plates,
over a quasi one-dimensional support of length $L$.
Rice grains are added to the left side, where a vertical bar
between the two plates forms a wall that keeps the grains inside
the system.
In contrast, the right side is open, and allows the exit of the grains 
out of the pile.
Starting with an empty system, the slow addition of grains makes the pile
to grow until the profile reaches the open boundary at the right.
After this transient time, the pile arrives to a quasi-stationary
state where the average slope fluctuates around a well defined 
angle of repose, and the influx of grains at the left equals
on average the outflux at the right exit.
At this point it has been shown by the Oslo group \cite{Frette96}
that the rice pile displays SOC if the shape of the grains is 
anisotropic enough to prevent the rolling of the grains down the slope,
suppressing the inertial effects and enhancing the dissipation by means
of the friction.
This result, apart of being the first unambiguous evidence of SOC
in granular media \cite{Frette96,Feder95} remarks the fact that
SOC is associated to strongly dissipative systems.

Christensen {\it et al.} \cite{Christensen96} 
have studied the transport 
properties of individual grains through the Oslo rice pile.
The transit time of tracer grains, defined as the time
necessary for a tracer to escape from the pile, was measured experimentally.
The results led to
a power-law distribution of transit times
for long times preceded by a flat region, i.e.,
\begin{equation}
    P_{tr}(T) \sim \left \{
    \begin{array}{ll}
            \mbox{constant for small $T$,}
    \\
            1/T^{\alpha} \ \ \mbox{ for large $T$},
    \end{array} \right.
\label{P_transit_exp}
\end{equation}
with $\alpha=2.4 \pm 0.2 $ \cite{Christensen96}.
Moreover, when the system size is varied, the distribution verifies
a finite size scaling,
\begin{equation}
     P_{tr}(T,L)=L^{- \nu'}
     F ( T/L^{ \nu} ),
\label{fss}
\end{equation}
with  $\nu=1.5 \pm 0.2$ and  $\nu'=1.4 \pm 0.2$ \cite{Christensen96}.
The fact of having $\nu \simeq \nu'$ follows from the normalization
condition.
From here, the scaling of the mean transit time with system size
was found to be "very anomalous":
\begin{equation}
    \langle T \rangle  \sim L^{\nu}.
\label{mean_transit_time}
\end{equation}
%

The goal of this Letter is to study the microscopic properties of the
transport of grains inside the rice pile, by means of a continuous
time random walk model \cite{Montroll}.
Comparing our theory with the experimental findings, 
we can give a form for the distribution of trapping times 
and the distribution of flights in the real system.
In addition we test our conclusions using a one dimensional
cellular automaton modeling the experiment \cite{Christensen96}
that connects transport with avalanche dynamics.

We are going to consider the motion of a single grain or particle 
through the profile as essentially one dimensional. 
This can be done because in the experiments the path of the particles
takes place between two points, it starts at the top of the pile, 
next to the left wall, and ends at the rightmost extreme of the support.
The rice grain remains at rest, trapped at position 
$x$ during a random time interval $t$
until some avalanche reaches it. 
At this point the grain performs
an instantaneous jump, or flight, of random length $l$,
after which it becomes trapped at $x+l$.
Then the dynamics of a particle is described in terms of a 
continuous time random walk, fully specified by the distribution of
trapping times $\psi(t)$ and by the distribution of flights $\phi(l)$.
To be precise $\psi(t)dt$ is the probability that the particle is trapped
at a given position a time between $t$ and $t+dt$,
whereas $\phi(l)dl $ gives the probability that the particle jumps 
a distance between $l$ and $l+dl$ during an avalanche.
Notice that we can assume $l>0$ always, since in the experiment no
mechanism allows the grains to climb the profile, always decreasing to
the right.
This will be a great simplification in the calculations
in comparison with models for diffusion in amorphous semiconductors
\cite{Schlesinger}.
In addition, the length of the flight $l$ will not be limited by 
the system size.
Both variables $t$ and $l$ are taken as independent 
random processes. 
The assumption of statistical independence has succeeded
in reproducing experimental results,
for instance in Ref. \cite{Morales}.
Finally note that with the hypothesis of instantaneous jumps
we are nothing else than fulfilling the usual condition for SOC,
that is to have a slowly driven system with two separated time scales,
where the motion of grains (or avalanches) happens at infinite velocity
in comparison with any external time scale.

The magnitude of interest 
is the distribution of transit times $P_{tr}(T,{\cal L})$,
where $P_{tr}(T,{\cal L})dT$ gives the
probability that the particle takes a time between $T$ and $T+dT$ to
travel from the origin to position $x={\cal L} \le L$.
In the context of stochastic processes the transit time is 
in fact the first passage time to level ${\cal L}$.
This distribution can be easily related with $p(x,t)$,
defined in such a way that $p(x,t)dx$ is the probability that 
at time $t$ a particle is in a position between $x$ and $x+dx$, 
the time being measured since the addition of the particle.
The probability of being at $x \le {\cal L}$ at time $t$
is equal to the probability of having a transit time $T>t$;
%
mathematically,
\begin{equation}
   \int_0^{\cal L} p(x,t)dx=1-\int_0^t P_{tr}(T,{\cal L})d T.
\label{pP}
\end{equation}
On the other hand $p(x,t)$ depends on the renewal density
$h(x,t)$, where $h(x,t)dx$ gives the probability of jump 
per unit time in a position between $x$ and $x+dx$ \cite{Cox}.
We can write close equations relating $p(x,t)$ and $h(x,t)$,
using $\psi(t)$ and $\phi(l)$:
\begin{eqnarray}
    \nonumber
    h(x,t)=[\mu \phi(x) +(1-\mu)\delta(x)] \psi(t) \\
    + \int_0^t \int_0^x h(x',\tau)\phi(x-x')\psi(t-\tau)dx'd\tau,
\label{h_and_p_1}    
    \\
    \nonumber
    p(x,t)= [\mu \phi(x) +(1-\mu)\delta(x)] \Psi(t)   \\
    + \int_0^t \int_0^x h(x',\tau)\phi(x-x')\Psi(t-\tau) d\tau dx',
\label{h_and_p_2}
\end{eqnarray}
where $\Psi(t) \equiv \int_t^{\infty} \psi(t')dt'$ 
gives the probability that the particle survives a time larger 
than $t$ trapped at any position.                                                      
$\mu$ is the probability that at $t=0$ the particle is moving.
Equations (\ref{pP})-
(\ref{h_and_p_2}) 
contain the solution to our problem, relating a measurable magnitude,
$P_{tr}(T,{\cal L})$ with the magnitudes that define the microscopic
dynamics, $\psi(t)$ and $\phi(l)$. 
Applying the Laplace transform, defined as
$ \hat f(\omega,s) \equiv \int_0^{\infty}dx \int_0^{\infty}dt \
     e^{-\omega x} e^{-s t} f(x,t)$,
the equations become linear and straightforwardly solvable.
In particular, $P_{tr}$ turns out to be
\begin{equation}
     \hat P_{tr}(s,\omega)=\frac{1}{\omega}
     \left [ 1- \frac{(1-\hat \psi(s))[ 1-\mu(1-\hat \phi(\omega))]}
     {1-\hat \psi(s) \hat \phi(\omega)} \right ].
\label{P_transit}
\end{equation}
The Laplace transform of the mean transit time 
$\langle T({\cal L}) \rangle $
can be easily obtained from $\hat P_{tr}(s,\omega)$ as
\begin{equation}
     \langle T(\omega) \rangle  = - 
     \left [ \frac{\partial}{\partial s}
      \hat P_{tr}(s,\omega) \right ]_{s=0} =
      \frac{1-\mu (1- \hat \phi)}{\omega (1-\hat \phi)}
      \langle t \rangle ,
\label{T_omega}
\end{equation}
where the existence of $\langle T(\omega) \rangle $ depends directly on the
existence of the first moment of $\psi(t)$, 
$\langle t \rangle  \equiv \int _0^{\infty}t \psi(t)dt = 
- (d \hat \psi/ d s)_{s=0}$.
As in the experiment $\langle T({\cal L}) \rangle $ was found to be finite,
we conclude that $\langle t \rangle $ exists as well.
The same reasoning for $\langle T^2({\cal L}) \rangle $, that was infinite in the 
experiment, implies that $\langle t^2 \rangle $ 
is not defined and then we assume
\begin{equation}
    \psi(t) \sim B/t ^{2+\beta}, 
    \ \mbox{ when } t \rightarrow \infty,
    \mbox{ with } 0 < \beta \le 1.
\label{psi}
\end{equation}
This means that in Laplace space \cite{Bleistein}
\begin{equation}
    \hat \psi(s) \sim 
    \ 1- \langle t \rangle s + B \Gamma (-1-\beta) s ^{1+\beta}
    \mbox{ when } s \rightarrow 0,
\label{psi_s}
\end{equation}
where $\Gamma(\cdot)$ is the Gamma function.
Substituting this expression into the equation for $\hat P_{tr}$
(\ref{P_transit}) and inverting the Laplace transform for $s$
we obtain
\begin{equation}
    \hat P_{tr}(t,\omega) \sim 
    \frac{1-\mu (1- \hat \phi)}{\omega (1-\hat \phi)}
    \frac{B}{t^{2+\beta}},
\label{P_transit_2}
\end{equation}
for $t \rightarrow \infty$.
Comparing (\ref{psi}) and (\ref{P_transit_2}) we see how the 
distribution of transit times, that is a macroscopic quantity,
is closely related with the distribution of trapping times,
a microscopic magnitude. In fact, both are long-tailed distributions
with the same exponent.
Linking with the experimental result (\ref{P_transit_exp})
we can conclude that the trapping times are power-law distributed
with an exponent
\begin{equation}
    2+\beta= \alpha.
\label{beta_alpha}
\end{equation}
Notice that the tail of the transit time distribution
only depends on the trapping time distribution,
and not on the distribution of jumps.
Alternatively, going back to Eq. (\ref{P_transit}) we can perform
first the long distance limit, i.e., $\omega \rightarrow 0 $.
If we consider that the jump distribution has mean value
but not second moment, that is, 
\begin{equation}
      \phi(l) \sim C/l^{2+\gamma}
      \ \mbox{when $l \rightarrow \infty$, with $0 < \gamma \le 1$},
\label{phi_l}
\end{equation}
then $\hat \phi(\omega)$ verifies an equation similar to 
(\ref{psi_s}), that substituting on (\ref{P_transit})
and inverting the Laplace transform gives
\begin{equation}
    \hat P_{tr}(s,{\cal L}) \sim \frac{C}{1+\gamma} \left 
    ( \mu + \frac{\hat \psi}{1-\hat \psi} \right ) 
    \frac{1}{{\cal L}^{1+\gamma}},
\label{P_transit_3}
\end{equation}
when $ {\cal L} \rightarrow \infty$.
This behavior will correspond to times "smaller" than ${\cal L}$.
For small times, the movement of the grains will be shallow
and the probability of a given transit time up to a position
${\cal L}$ will be independent on system size; 
in other words, Eq. (\ref{P_transit_3}) will not depend on $L$.
If we make ${\cal L}=L$, this means that the scaling with
system size is the same than with position.
Denoting the latter by subindices $\nu_x$ and $\nu_x'$, i.e., 
$P_{tr}(T,{\cal L})={\cal L}^{-\nu_x'}{\cal F}(T/{\cal L}^{ \nu_x})$,
this implies that $\nu'=\nu'_x$,
and comparing with the scaling found in the experiment (\ref{fss})
we have
\begin{equation}
    1+\gamma=\nu'.
\label{gamma_nu}
\end{equation}
One can consider other asymptotic forms for $\phi(l)$,
but it is only the one given by Eq. (\ref{phi_l}) that reproduces an
exponent $\nu'$ between 1 and 2.
From here we can deduce that in the rice pile the scaling 
of $P_{tr}$ given by (\ref{fss}) means that the distribution
of jumps has a finite mean $\langle l \rangle$
but an infinite variance.
This kind of distributions correspond to L\'evy flights
\cite{Klafter}, and give rise to a superdiffusive behavior,
as one can verify by finding
$\hat p(\omega,s)$ and from here obtain 
(similarly as in Eq. (\ref{T_omega})) that
$\langle x(t) \rangle \sim t \langle l \rangle /\langle t \rangle $
and $\langle x^2(t) \rangle \propto \langle l^2 \rangle =\infty$.

Moreover, if the limit $s \rightarrow 0$
is performed in Eq. (\ref{P_transit_3}), 
one obtains the behavior for large times
but "smaller" than ${\cal L}$,
that turns out to be independent on $t$:
\begin{equation}
    P_{tr}(t,{\cal L}) \sim \frac{C}{(1+\gamma)\langle t \rangle}
    \frac{1}{{\cal L}^{1+\gamma}},  
    \ \mbox{ when } t < \frac{\langle t \rangle}{\langle l \rangle}
    {\cal L}.
\label{scaling_1}   
\end{equation}
This corresponds to the flat region observed in the
transit time distribution before the power-law decay.
In fact, the appearance of a plateau in $P_{tr}$
is an indication of the existence of $\langle t \rangle$.

Now that the asymptotic form of $\phi(l)$ is known,
one can go back to Eqs. (\ref{P_transit_2}) and (\ref{T_omega})
to perform the long distance limit in order to obtain the scaling
of $P_{tr}$ with ${\cal L}$ for large times and
the scaling of the mean transit time:
\begin{equation}
     P_{tr}(t,{\cal L}) \sim \frac{B}{\langle l \rangle}
     \frac{{\cal L}}{t^{2+\beta}} 
     \ \mbox{ when } t \gg \frac{\langle t \rangle}{\langle l \rangle}
     {\cal L},
\label{scaling_2}   
\end{equation}
\begin{equation}
     \langle T({\cal L}) \rangle \sim
     {\cal L}{\langle t \rangle}/{\langle l \rangle}
     \ \mbox{ when } {\cal L} \rightarrow \infty.
\label{mean_transit_time_2}   
\end{equation}
From here one obtains that $\nu_x=1$, in contrast with the
value of $\nu_x'=1+\gamma$.
The reason to have $\nu_x \ne \nu_x'$ is simple: 
the model does not show finite size scaling for all
$T$ and ${\cal L}$.
Indeed we have found scaling only
for a region  $T<{\cal L}\langle t \rangle /\langle l \rangle $
and $T \gg {\cal L}\langle t \rangle /\langle l \rangle $.
By using numerical simulations we will see that this behavior
is right and one can expect in the experiment
$\langle T ({\cal L})\rangle \sim {\cal L}$ as well.
If we impose the continuity of $P_{tr}$ at the crossover
point $T_c \sim {\cal L}$ we obtain
\begin{equation}
     \beta=\gamma,
\label{beta_gamma}
\end{equation}
that is, $\psi(t)$ and $\phi(l)$ must have the same
power-law tail.
Employing also Eqs. (\ref{beta_alpha}) and (\ref{gamma_nu})
we get 
%
%
%
\begin{equation}
    \alpha=1+ \nu'.
\label{alpha_nu}
\end{equation}
This equation relates the exponent of the power-law tail of
$P_{tr}$ with its scaling with system size,
and it is well fulfilled by the experimental values,
see Eqs. (\ref{P_transit_exp}) and (\ref{fss}).

It would be difficult to design an experiment to measure
the distributions of trapping times and jumps, 
that are microscopic magnitudes, although in granular materials
microscopic and macroscopic scales are not so well separated
as in the usual states of matter \cite{Jaeger}.
As an alternative to support our predictions
we use the cellular automaton model
introduced in Ref. \cite{Christensen96}, 
which was found to reproduce the transport properties
of grains quite well.
We believe that similar results can be obtained
for similar models \cite{Amaral}.
On a one-dimensional lattice, from $x=1$ to $L$,
an integer variable $h_x$ gives the height of the pile
at position $x$.
Defining the local slope at $x$ as $z_x \equiv h_x-h_{x+1}$
the dynamics of the model is fully determined by the following
rules:
if $z_x \le z_x^c \ \forall x  \Rightarrow 
         z_1 \rightarrow z_1+1$
         (a grain is added); 
if  $z_x > z_x^c \mbox{ and } x<L \Rightarrow
         z_{x-1} \rightarrow z_{x-1}+1,
         z_{x} \rightarrow z_{x}-2,
         z_{x+1} \rightarrow z_{x+1}+1, \mbox{ and } 
         z_{x}^c \rightarrow rand(1,2)$;
if $z_L > z_L^c \Rightarrow  
         z_{L-1} \rightarrow z_{L-1}+1,
         z_{L} \rightarrow z_{L}-1, \mbox{ and } 
         z_{L}^c \rightarrow rand(1,2)$; 
%
where all the sites have to be updated in parallel
and $rand(1,2)$ means 1 or 2 at random, with equal probability.
The external input of grains at $x=1$ sets the time unit.

%
\begin{figure}
\epsfxsize=2.2truein 
\hskip 0.15truein\epsffile{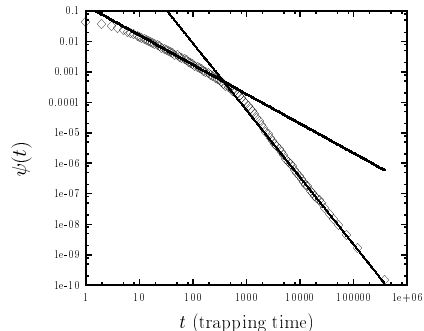} 
\caption{
Trapping time distribution $\psi(t)$ in a system of size $L=400$.
The two straight lines are power laws with exponents
$-.97 \pm 0.05$ and $-2.20 \pm 0.05$.
\label{waiting}
}
\end{figure}
\begin{figure}
\epsfxsize=2.2truein 
\hskip 0.15truein\epsffile{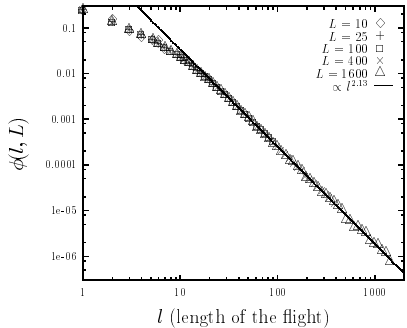} 
\caption{
Distribution of flight lengths $\phi(l,L)$ for different system sizes
starting from position $x=2$.
The length of the flight does not depend on $L$.
The power-law behavior for large distances is characterized
by an exponent $-2.13 \pm 0.05$.
\label{flight_L}
}
\end{figure}

The results of Ref. \cite{Christensen96} show that Eqs.
(\ref{P_transit_exp})-(\ref{mean_transit_time})
are still valid, but the exponents are determined with more accuracy.
We reanalyze these results to obtain
$ \alpha=2.21 \pm 0.05, 
\nu=1.25 \pm 0.10, \mbox{ and } \nu'=1.25 \pm 0.10$,
in concordance with (\ref{alpha_nu}).
The distribution of trapping times can be obtained from
simulations as the number of trappings of a given duration
divided by the total number of trappings. 
The results are displayed in Fig. \ref{waiting}.
Indeed we obtain a power-law distribution for long times,
where the exponent turns out to be
$     2+\beta=2.20 \pm 0.05$,
in very good agreement with our prediction (\ref{beta_alpha})
if we compare with the
independent measure of $\alpha$.
One can also measure the distribution of flight lengths.
The behavior for long distances corresponds indeed
to a power law, whose exponent is
$     2+\gamma=2.13 \pm 0.05$,
see Fig. \ref{flight_L}, where we restrict the measure
to flights starting at a fixed position.
Observe from here that $\nu'$, $\beta$, and $\gamma$ 
are compatible with (\ref{gamma_nu}) and (\ref{beta_gamma}).
The simulations allow one also to study the transit time
to reach a certain position $x={\cal L}$ smaller
than the system size $L$. 
Keeping fixed $L$ the exponents of the scaling of 
$P_{tr}$ with ${\cal L}$ are
$     \nu_x=1.0 \pm 0.1
     \ \mbox{ and } \
     \nu_x'=1.2 \pm 0.1 $,   
as was predicted by Eqs. (\ref{scaling_1}) and (\ref{scaling_2})
whereas the power-law exponent $\alpha=2.18 \pm 0.05$
is in good accord with (\ref{alpha_nu}).
One can additionally measure the mean transit time as a
function of ${\cal L}$ to verify that Eq. (\ref{mean_transit_time_2})
is fulfilled.
\begin{figure}
\epsfxsize=2.2truein 
\hskip 0.15truein\epsffile{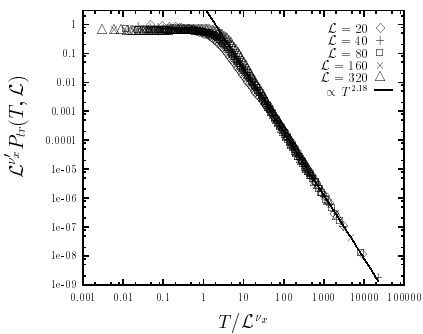} 
\caption{
Scaled distributions of transit times from $x=1$ to $x={\cal L}$
in a system of size $L=400$ for different values of ${\cal L}$.
The scaling is done for $T \gg {\cal L}$ and $T < {\cal L}$,
resulting that the exponents are $\nu_x=1.0$ and $\nu_x'=1.2$,
whereas the power-law  exponent is $2.18 \pm 0.05$.
Notice that the region with $T$ slightly larger than ${\cal L}$
does not scale well.
\label{transit_x}
}
\end{figure}
Moreover we can derive an additional scaling relation for the model.
In Ref. \cite{Christensen96} it was found that
$\chi=\nu-1$, being $\chi$ the roughness exponent of the 
profile of the pile, that is, the fluctuations of the
profile scale as $L^{\chi}$. 
On the other hand, in Ref. \cite{Paczuski} it was argued
that $D=2+\chi$, with $D$ the fractal dimension of the avalanches,
i.e., the size of the avalanches scales as $L^D$.
As $\nu \simeq \nu'$, combining these relations with (\ref{alpha_nu})
one gets
\begin{equation}
     \alpha=D
\end{equation}
Taking $D=2.23 \pm 0.03$ \cite{Paczuski} this last result
is in a fair 
agreement with the measured value of $\alpha$.

In summary, from analytical results and computer simulations
we present a coherent scenario for the transport in a
self-organized critical granular system.
The scale invariance of the process is associated to long-tailed
trapping time distributions and L\'evy flights of the grains.

It is a pleasure to acknowledge K. Christensen and J.M. Porr\`a
for discussions and suggestions.
M.B. is partially supported by CICyT's research program \#PB96-0188,
whereas A.C. was supported by \#PB94-0897
and a grant of the Spanish MEC.


\end{document}